# MECHANISMS OF DIFFUSIONAL NUCLEATION OF NANOCRYSTALS AND THEIR SELF-ASSEMBLY INTO UNIFORM COLLOIDS

Vladimir Privman

Center for Advanced Materials Processing, Clarkson University, Potsdam, NY 13699

www.clarkson.edu/Privman

## ABSTRACT

We survey our research on modeling the mechanisms of control of uniformity in growth of nanosize and colloid particles. The former are produced as nanocrystals, by burst-nucleation from solution. The latter, colloid-size particles, are formed by self-assembly (aggregation) of the nanocrystals. In the colloid particle synthesis, the two dynamical processes are coupled, and both are governed by diffusional transport of the respective building blocks (monomers). The interrelation of the two processes allows for synthesis of narrow-size-distribution colloid dispersions which are of importance in many applications.

We first review a mathematical model of diffusive cluster growth by capture of monomer "singlets." Burst nucleation of nanoparticles in solution is then analyzed. Finally, we couple it to the secondary process of aggregation of nanoparticles to form colloids, and we discuss various aspects of the modeling of particle size distribution, as well as other features of the processes considered.

---

---

# 1. INTRODUCTION

There have been new interesting recent trends in colloid and nanoparticle science that require novel theoretical developments. Specifically, there has been a drive to devise controlled synthesis approaches for fine particles starting from colloids (micron and sub-micron size particles) down to nanoparticles of dimensions of 0.01 µm, or 10 nm, and smaller.

Quantification of the kinetics of nucleation, growth, aggregation, and surface interactions of nanoparticles require new experimental probes, but also new theoretical techniques. Here we survey our recent results[1-9] on quantitative modeling of the process of burst nucleation and diffusive growth of nanoparticles in solution, as well as on the accompanying secondary process of diffusive aggregation of the resulting, typically crystalline nanoparticles to form uniform polycrystalline colloids.

Modeling of nanoparticle and colloid formation in solution, is a active field with many open problems and experimental as well as theoretical challenges. We have developed quantitatively successful modeling schemes[1,3,5-8] to explain the narrow size distributions observed for properly selected experimental conditions in synthesis of "monodispersed" colloidal particles of various compositions. More recently, we have addressed quantitatively[9] the particle size distribution in the model of burst nucleation, which, however, in its "classical" form is expected to be at best only approximately valid for real nanoparticle synthesis.

In Section 2, we offer a general discussion of the particle size selection mechanisms, as well as of a mathematical mechanism for diffusive growth by capture of monomers. Our results for burst nucleation of nanoparticles are outlined in Section 3, in which we also address the limitations of the model. When burst nucleation is accompanied by the secondary process of nanoparticle aggregation, then self-assembly of uniform particles of colloid dimensions results. This phenomenon is surveyed in Section 4. Finally, in Section 5 we discuss some additional developments and open problems, the latter specifically the shape selection and shape distribution in uniform fine particle synthesis.

# 2. SIZE SELECTION IN FINE PARTICLE SYNTHESIS

The concept of "monodispersed" particles in colloid applications usually means distributions of relative width 6-12%. An important conceptual issue involves extensions to nanosize particles. What do we mean by "monodispersed" at the nanoscale? It is quite likely that for most truly large-molecule-dimension nanotechnology applications, uniform size (and shape) really means "atomically identical." This is particularly true for future electronic devices. For many other applications, requirements for uniformity will also be quite strict.

Therefore, methods of controlling size and shape distributions, which found numerous applications for colloids, will be even more important for nanotechnology. Here we consider situations with monomer "building blocks" for particle formation, as well as particles themselves, transported by diffusion in solution. The singlet (monomer) building blocks in nanoparticle synthesis in solution are atomic-size solute species (atoms, ions, molecules), whereas for colloid synthesis they are the (nanosize) primary particles. In the latter case, the supply of singlets is "naturally" controlled by the parameters of their burst-nucleation process. However, in principle the solutes can be also added/mixed in externally.

A typical particle size distribution of interest is illustrated in Figure 1 (see Page 22). We comment that mechanisms such as cluster-cluster aggregation, or cluster ripening due to exchange of monomers, while making the size distribution slowly grow, will also broaden it. They cannot lead to size selection. Generally, most growth/coarsening mechanisms that involve diffusional transport broaden the distribution because larger particles have larger collection area for capturing "building blocks," as well as, e.g., for spherical particles, less surface curvature, which implies generally slightly better binding of monomers (less detachment on average).

Narrow particle size selection can be achieved by several techniques. The simplest is to actually block the growth of the "right side" of the peak, see Figure 1 (Page 22), by "caging" the particles. An example could be nanoparticles grown inside nanoporous structures or objects. We do not consider this technique,[10] because it falls outside the theoretical and experimental framework of the topical coverage of this article.

Another approach involves dynamical processes that "erode" the left side of the "pedestal" of the peak in the distribution, fast enough as compared to the peak broadening by coarsening processes, to maintain narrow distribution. The burst-nucleation process analyzed in Section 3, falls in this category. Unfortunately, other coarsening processes can eventually broaden the distribution after the initial nucleation burst. We will return to these issues in Section 3.

Another approach to obtain particle size distributions narrow on a *relative* scale, involves fast supply of monomers, of concentration $C(t)$, see Figure 1 (Page 22). The monomers "feed" the peak, thus pushing it to larger average sizes, and the process can be fast enough not to overly broaden the distribution (on a relative scale) and, with proper time-dependent $C(t)$, not to generate too large a "shoulder" of small clusters.

It is therefore quite natural to consider the time dependence of the singlet (monomer) availability, and its impact on the size distribution of the products. Specifically, for nanosize particle preparation, there has been interest[11,12] in stepwise processes: after achieving the initial nanoparticle distribution, batches of singlets are added to induce further growth.

Let $N_s(t)$ denote the volume density of particles, consisting of $s$ singlets, at time $t$. We are interested in the situation illustrated in Figure 1 (see Page 22), when the particle size distribution evolves in time with a peak eventually present at some relatively large $s$ values. For convenience, let us denote the singlet concentration by

$$C(t) \equiv N_1(t). \tag{2.1}$$

The singlets can be supplied as a batch, several batches, or at the rate $\rho(t)$, per unit volume. They are consumed by the processes involving the production of small clusters, in the "shoulder" in Figure 1 (see Page 22). They are also consumed by the growing large clusters in the peak.

There are two issues to consider in such growth: how is the peak created in the first place, and how to grow it without much broadening. In this section, we primarily address the latter issue. Regarding the former, for nanoparticle synthesis the main mechanism of the early formation of the peak is by burst nucleation, when nuclei of sizes larger than the critical size form by growing over the nucleation barrier. Of course, *seeding* is another way of initiating the peaked size distribution both for colloid and nanoparticle growth. For colloid synthesis without seeding, the initial peak formation is more subtle and could actually be a result of few-singlet cluster-cluster aggregation at the early growth stages, as further commented on at the end of Section 5.

Here we consider growth dominated by irreversible capture of singlets by the larger growing aggregates. Thus, we use the rate equations (master equations), with $\Gamma_s$ denoting the rate constants for singlet capture by the $s \geq 1$ aggregates,

$$\frac{dN_s}{dt} = (\Gamma_{s-1}N_{s-1} - \Gamma_s N_s)C, \qquad \text{for } s > 2, \tag{2.2}$$

$$\frac{dN_2}{dt} = (\frac{1}{2}\Gamma_1 C - \Gamma_2 N_2)C, \tag{2.3}$$

and

$$\frac{dC}{dt} = \rho - \sum_{s=2}^{\infty} s \frac{dN_s}{dt} = \rho - \Gamma_1 C^2 - C \sum_{s=2}^{\infty} \Gamma_s N_s \,. \tag{2.4}$$

Let us point out that the assumption that the *only* process involving the $s > 1$ aggregates is that of capturing singlets at the rate proportional to the concentration of the latter, $\Gamma_s C$, is drastic but commonly used in the literature.[1,5–6,13–15] We will consider elaborations, see Section 5. As mentioned, more complex processes, such as cluster-cluster aggregation,[16,17] detachment[2,4] and exchange of singlets (ripening), etc., also contribute to particle growth. However, in uniform colloid synthesis they are typically much slower than the singlet-consumption growth. In addition, they broaden the particle size distribution.

Another important approximation involved in writing (2.2)-(2.4) is that of ignoring particle shape distribution and particle morphology. We avoid this issue, which is not well understood, by assuming that the growing aggregates rapidly restructure into compact bulk-like particles, of an approximately fixed shape, typically, but not always, spherical. This has been experimentally observed in uniform colloid synthesis.[1,18–23] Without such restructuring, the aggregates would be fractal.[17,24] We further comment on the shape selection issue as an unsolved problem, in Section 5.

For nanosize particle synthesis, the assumption in the above summary that should be questioned is that of ignoring singlet detachment (and more generally "embryo" breakup) for the particles in the shoulder in Figure 1 (see Page 22). Indeed, unlike colloid growth, which is fast and irreversible for all $s$ in solution synthesis processes, the nanosize particle growth will be typically held back by a nucleation barrier.[1,6,9,11] During the late stage growth, that follows the initial nucleation burst,[9,25,26] the barrier can be quite high. The distribution in the shoulder will approach the equilibrium Boltzmann form, governed by the excess free energy of the aggregate formation. It is interesting to note that this fast equilibration means that the singlets "stored" in the small, "shoulder" aggregates will be ultimately available for consumption by larger aggregates in the peak. The resulting "burst nucleation" is analyzed in Section 3.

Here we thus focus on the situations for which the assumptions leading to (2.2)-(2.4) apply: "minimal" models of colloid growth and certain stepwise nanoparticle growth processes. If the singlets are supplied/available constantly, then the distribution, both for colloids and nanoparticles, will develop a large shoulder at small aggregates, with no pronounced peak at $s \gg 1$. If the supply is limited, then only small aggregates will be formed. Our key recent discovery in studies of colloid synthesis[1,6] has been that there exist "protocols" of singlet availability, at the rate $\rho(t)$ which is a slowly decaying, sometimes rather complicated function of time, that yield peaked size distributions at large times. Furthermore, the primary (nanocrystal nucleation) process in uniform polycrystalline colloid synthesis, naturally "feeds" the secondary process (of nanoparticles aggregation) just at a rate like this.

Solution of the rate (master) equations (2.2)-(2.4) requires numerical approaches and is not particularly illuminating as to the nature of the particle growth. Therefore, to explore the nature of the peak growth, in the rest of this section we will introduce several additional assumptions which will allow us to go a long way in simplifying the problem in closed analytical form. The main idea is that, once the peak is formed after some transient time or by seeding, the particles in the peak are the main consumers of the available singlets.

This assumes that the singlet concentration is controlled by adding them externally.[6,11,12] For nanoparticles, the addition should be at such a rate that the nucleation barrier remains high. The shoulder will then adjust to assume an approximately equilibrium shape, but the production of new larger, supercritical aggregates will be negligible. For colloid growth, the shoulder will also evolve, with new particles generated. However, if the number of larger aggregates is already significant, they will dominate the consumption of singlets.

Thus, to understand how a well-developed peak can evolve while remaining relatively narrow, let us entirely inhibit generation of new small aggregates, by setting

$$\Gamma_1 \to 0, \tag{2.5}$$

for the rest of our derivation in this section, which is an approximation. Furthermore, we will assume that $s$ is a continuous variable, since we are interested here in $s \gg 1$, and that it varies in the range $0 \leq s < \infty$.

For calculations assuming singlet transport by diffusion, one can take the large-$s$ Smoluchowski expression for the rates,[2,27,28]

$$\Gamma_{s \gg 1} = \Upsilon s^{1/3}, \tag{2.6}$$

where $\Upsilon$ is a known constant. Note that $\Gamma_{s \gg 1}$ is proportional to the aggregate linear dimension (which yields the factor $s^{1/3}$) times the singlet diffusion constant. Our results in this section actually apply for general $\Gamma_s$. *A reader not interested in detailed mathematical derivations could skip the following steps and go directly to the paragraph containing the last equation, (2.21), of the present section.*

Our last approximation is introduced while deriving the continuous-$s$ form of (2.2): we retain only the leading $s$ derivative, ignoring here the "diffusive" second-derivative term (this will be remedied later, in Section 3). The consequences of this approximation, already used in the literature,[6,13] will be discussed later. Thus, we replace (2.2) by

$$\frac{\partial N(s,t)}{\partial t} = -C(t) \frac{\partial}{\partial s} [\Gamma(s) N(s,t)], \tag{2.7}$$

with (2.4) replaced by

$$\frac{dC(t)}{dt} = \rho(t) - C(t) \int_0^\infty ds [\Gamma(s) N(s,t)]. \tag{2.8}$$

Let us now define the variable

$$\tau(t) = \int_0^t dt' C(t') \geq 0, \tag{2.9}$$

and then introduce the function $u(s,\tau)$ via the relation

$$\tau = \int_u^s \frac{ds'}{\Gamma(s')}. \tag{2.10}$$

We point out that usually $\Gamma(s) > 0$, and the lower limit of integration can be taken to zero. The asymptotic rate expression (2.6) does vanish at $s = 0$ because of our cavalier treatment of the small-$s$ behavior. However, the integral happens to converge, so no additional care is needed. We can safely define the quantity $s_{\min}(\tau)$ via

$$\tau = \int_0^{s_{\min}} \frac{ds'}{\Gamma(s')}. \tag{2.11}$$

As $u$ is increased from zero to infinity, the corresponding $s(u,\tau)$, for fixed $\tau$, increases from $s_{\min}(\tau)$ to infinity.

Next, we notice that the relation between the differentials implied by (2.10), namely,

$$d\tau = \frac{ds}{\Gamma(s)} - \frac{du}{\Gamma(u)}, \tag{2.12}$$

allows us to calculate various partial derivatives in terms of $\Gamma(s)$ and $\Gamma(u) = \Gamma\big(u(s,\tau(t))\big)$. This, in turn, allows one to verify, by a cumbersome calculation not reproduced here, that (2.7) is solved by

$$N(s,t) = \frac{\Gamma\big(u(s,\tau(t))\big)}{\Gamma(s)} N\big(u(s,\tau(t)),0\big), \qquad \text{for } s \geq s_{\min}\big(\tau(t)\big), \tag{2.13}$$

and

$$N(s,t) = 0, \qquad \text{for } 0 \leq s \leq s_{\min}\big(\tau(t)\big), \tag{2.14}$$

where the discontinuity at $s_{\min}\big(\tau(t)\big)$ is possible if the initial distribution at time zero, $N(s,0)$, is nonzero at $s = 0$. Actually, within the present approximation of ignoring the effects of the details of the size distribution for small $s$, we could as well set $N(0,0) = 0$.

Let us summarize the above observations by emphasizing that we consider a particle size distribution which at time $t = 0$ already has a well-developed significant peak at large cluster sizes. Relations (2.13)-(2.14) will provide an approximate description of further evolution of this peak with time, due to supply of singlets at the rate $\rho(t)$. The form of the distribution at small particle sizes plays no role in the derivation.

In fact, neglecting the second-derivative in $s$, "diffusive" term in writing (2.7), leads to certain artificial features. Specifically, sharp corners and discontinuities of the initial distribution (as well as its derivatives, etc.) will not be smoothed out. The fact that the initial distribution is only meaningful for $s \geq 0$ translates into the sharp cutoff at $s_{\min}$ for times $t > 0$. Had we included the diffusive term, the distribution would extend smoothly to $s = 0$ for all times. However, no closed-form analytical solution would be available.

While this lack of smoothness is probably not important for a semi-quantitative evaluation of the size distribution, one aspect should be emphasized as critical: if the initial distribution is already very sharp, then the neglect of the diffusive term in our expressions may result in underestimating the *width* of the evolving peak.

To complete the description of the particle size distribution within the non-diffusive approximation, we have to discuss the estimation of the function $\tau(t)$. Relations (2.8)-(2.9) can be rewritten, using (2.13), as a system of coupled differential equations for two unknown functions $\tau(t)$ and $C(t)$, with $\tau(0) = 0$, and $C(0)$ externally controlled,

$$\frac{d\tau}{dt} = C(t), \tag{2.15}$$

and

$$\frac{dC}{dt} = \rho(t) - C(t)F(\tau), \tag{2.16}$$

where

$$F(\tau) = \int_{s_{\min}(\tau)}^{\infty} ds\left[\Gamma(u(s,\tau))N(u(s,\tau),0)\right]. \tag{2.17}$$

These equations are easily programmed for numerical evaluation, especially if the function $F(\tau)$ is calculable analytically, so that numerical integration can be avoided. The latter is likely for the power-law rate in (2.6), provided the initial distribution $N(s,0)$ is not too complicated.

Within the approximation developed here, the number of particles larger than singlet, $M$, obviously remains constant,

$$M = \int_{s_{\min}(t)}^{\infty} ds N(s,t) = \int_0^{\infty} ds N(s,0). \tag{2.18}$$

The change in the average size of the particles larger than singlet,

$$\langle s \rangle_t = \frac{1}{M} \int_{s_{\min}(t)}^{\infty} ds\left[sN(s,t)\right], \tag{2.19}$$

can be evaluated directly from $C(t)$,

$$\langle s \rangle_t = \langle s \rangle_0 + \frac{1}{M}[C(0) - C(t) + \int_0^t dt' \rho(t')]. \tag{2.20}$$

Furthermore, consideration of the increment relations following from (2.12), suggests that the width of the peak, $W_t$, where the subscript denotes the time variable, grows according to

$$W_t \approx \frac{\Gamma(\langle s \rangle_t)}{\Gamma(u(\langle s \rangle_t, \tau(t)))} W_0 > W_0. \tag{2.21}$$

The inequality follows from the definition (2.10), assuming that for large $s$, $\Gamma(s) > 0$ is an increasing function. This excludes an important case of constant $\Gamma$, appropriate for certain models of polymerization. In that case, however, the discrete equations (2.2)-(2.4) can be analyzed directly in great detail,[14,15] so that the above mathematical formulation is not needed.

In connection with (2.21), the reader must be cautioned that additional broadening will result from the second-derivative "diffusive" term neglected in our continuous-$s$ equations. The model with the diffusive term included, requires serious numerical efforts, as does the original, discrete-$s$ model; however, see Section 3 for some explicit expressions.

In summary, with the reservations regarding the width (under)estimates, numerical calculation of the functions $\tau(t)$ and $C(t)$, via (2.15)-(2.17), goes a long way in estimating various properties of the growing, peaked size distribution.

Even at the level of the approximations leading to (2.21), it is obvious that the size distribution *never actually narrows in absolute terms*. Specifically, experimentally realized monodispersed particle synthesis procedures in solution, in the colloid domain, actually yield small *relative* peak width, $W_t / \langle s \rangle_t$, by utilizing fast increase in $\langle s \rangle_t$ via consumption of singlets, on the time scales too short for the "diffusive" broadening to set in.

## 3. BURST NUCLEATION

The model of burst nucleation[9,25,26] treats the growth of nanosize particles, typically, crystals, consisting of $n$ monomers (we will reserve $s$ for the count of singlets in growing colloids, see Sections 4-5), in the peak in the same way as described in the preceding section. These larger particles, with $n > n_c$, where $n_c$ is the critical cluster size, irreversibly capture atomic/molecular size "monomer" diffusing solutes. However, the dynamics in the "shoulder," for $n < n_c$, compare Figures 1 and 2 (see Pages 22-23), is no longer ignored. Rather, these subcritical embryos are assumed instantaneously re-thermalized.

We consider a supersaturated solution with time-dependent monomer concentration $c(t)$. Thermal fluctuations cause formation of aggregates (embryos), controlled by the free-energy barrier imposed by the surface free energy. Of course, the true dynamics of the few-atom clusters involves complicated transitions between embryos of various sizes, shapes, as well as internal restructuring. These processes are presently not well understood.

Therefore, it is usually assumed that the dynamics of embryos is very fast, and their sizes are approximately thermally distributed. This distribution can be modeled by a Gibbs-like form[1,5] of the free energy of an *n*-monomer embryo,

$$\Delta G(n,c) = -(n-1)kT \ln(c/c_0) + 4\pi a^2 \left( n^{2/3} - 1 \right) \sigma , \qquad (3.1)$$

where $k$ is Boltzmann constant, $T$ is the temperature, $c_0$ is the equilibrium concentration of monomers, and $\sigma$ is the effective surface tension. The first term is the "bulk" free-energy contribution. It is derived from the entropy of mixing of noninteracting solutes and is negative for $c > c_0$, therefore favoring larger clusters. The second, positive term represents the surface free-energy cost, proportional to the surface area, $\sim n^{2/3}$. The effective solute radius, $a$, is defined so that the radius of an $n$-solute embryo is $an^{1/3}$. It can be estimated by requiring that $4\pi a^3/3$ equal the "unit cell" volume per monomer (including the surrounding void volume) in the bulk material.

As in most treatments of homogeneous nucleation, we assume that the distribution of embryo shapes can be neglected: an "average" cluster is assumed spherical in the calculation of its surface area and the monomer transport rate to it. We note that even the surface tension of spherical particles varies with their size. This effect, as well as any geometrical factors that might be needed because real clusters are not precisely spherical, is neglected. The effective surface tension of nanoparticles is only partially understood at present.[29] Thus, $\sigma$ can be either assumed[1,5,7,8] close to $\sigma_{\text{bulk}}$, or fitted as an adjustable parameter.

The free energy (3.1) increases with $n$ until it reaches the "peak of the nucleation barrier" at $n_c$,

$$n_c(c) = \left[\frac{8\pi a^2 \sigma}{3kT \ln(c/c_0)}\right]^3 = \left[\frac{2A}{3\ln(c/c_0)}\right]^3, \tag{3.2}$$

where $A \equiv 4\pi a^2 \sigma/kT$. For $n > n_c$, the free energy decreases with $n$, and it is then assumed that the kinetics becomes irreversible and is no longer controlled by $\Delta G$.

The specific property of burst nucleation is that the barrier, and $n_c$, markedly depend on the monomer concentration, $c$, which leads to a significant suppression of nucleation after the initial burst, during which $c/c_0$ decreases by several orders of magnitude: from its initial value $c(0)/c_0 \gg 1$ to it asymptotic large-time value 1.

The large-time form of the particle size distribution in burst nucleation is shown in Figure 2 (see Page 23). Specifically, the embryonic matter below $n_c$ is thermalized on time scales much faster than those of other dynamical processes, so that the concentration of embryos, with sizes in $dn$, is given by $P(n,t)dn$, with the particle size distribution

$$P(n < n_c, t) = c(t) \exp\left[\frac{-\Delta G(n, c(t))}{kT}\right], \tag{3.3}$$

where $n_c = n_c(c(t))$.

The rate of production of supercritical clusters, to be denoted by $\rho(t)$ for use in Section 4, is then expressed[1] as

$$\rho(t) = K_{n_c} c P(n_c, t) = K_{n_c} c^2 \exp\left[\frac{-\Delta G(n_c, c)}{kT}\right], \tag{3.4}$$

where $K_n = 4\pi a n^{1/3} D$ is the Smoluchowski expression[2,27,28] for the rate of intake of diffusing solutes by spherical particles. We already encountered this rate in (2.6), and we note that we use the large-$n$ form for supercritical clusters, $n \geq n_c \gg 1$. Here $D$ is the diffusion coefficient for monomers in a solution with viscosity $\eta$; up to geometrical factors, $D$ can be estimated as $\sim kT/6\pi\eta a$.

Although real clusters undergo both attachment and detachment of monomers (with detachment still present at sizes above $n_c$), we model the expected rapid growth of the supercritical, $n > n_c$, clusters within the approximation of irreversible capture of diffusing monomers (no detachment), using the master equation

$$\frac{\partial P(n,t)}{\partial t} = (c(t) - c_0)(K_{n-1}P(n-1,t) - K_n P(n,t)). \tag{3.5}$$

Comparing to (2.2), the difference $c(t) - c_0$ is used in place of $c(t)$ to ensure that the growth of clusters stops when the equilibrium concentration $c_0$ is reached. In actuality, the variation of surface tension with particle radius mentioned above is accompanied by a variation of the effective equilibrium concentration with radius, which gives rise to Ostwald ripening.[30] This, as well as other possible coarsening processes, such as cluster-cluster aggregation,[16,17,31] are neglected here because burst nucleation is expected[1,9] to be a much faster process. However, for large times such coarsening processes will gradually widen the particle distributions seen in experiment and slow down the growth of the particle size, which, as will be argued shortly, for burst nucleation alone is well characterized by the function $n_c(t)$ schematically shown in Figure 2 (see Page 23).

We further comment that in addition to growth (shrinkage) by attachment (detachment) of monomers, clusters of all sizes can undergo internal restructuring, a complex phenomenon the modeling of which for nanoscale clusters is only in its early stages.[32,33] Without such restructuring, the clusters would grow according to diffusion-limited aggregation or similar processes and could be fractals,[16,17] while observations of the density and X-ray diffraction data of colloidal particles aggregated from burst-nucleated nanocrystalline subunits indicate that their polycrystalline structure has the *density of the bulk*.[1,34] There is primarily experimental, but also modeling evidence,[1,4,5,7,8] that for larger clusters such restructuring leads to compact particles with smooth surfaces, which then grow largely irreversibly.

The "right side" of the supercritical distribution, see Figure 2 (Page 23), grows towards larger clusters by capturing monomers, but, at the same time, its "left side" is eroded by the thermalized subcritical distribution which extends up to $n_c(t)$, which is a monotonically increasing function of time. The form of the supercritical distribution depends on the initial conditions. As will be demonstrated shortly, at large times it will eventually have its maximum at $n = n_c$, and will take on the form of a truncated Gaussian. This is illustrated in Figure 2 (see Page 23), where the peak of the full Gaussian curve (not shown) is actually to the left of $n_c$.

Numerical results for time-dependent distributions and for several initial conditions, presented in,[9] were obtained by a novel efficient numerical integration scheme which is not reviewed here. In what follows, we concentrate on the derivation of analytical results for large times. We note that one must be consistent, in both the asymptotic and numerical treatments, with the conventions for relating the discrete-$n$ quantities, such as the monomer concentration $c(t)$, to the values of the continuous distributions. We have chosen the simple convention $c(t) = P(1,t)$, rather than, e.g., a convention to treat the monomer concentration $c(t)$ separately of the rest of the distribution, as was done in Section 2. Then the conservation of matter is expressed as the quantity

$$\int_1^{n_c} n\, c(t) \exp\left[\frac{-\Delta G(n, c(t))}{kT}\right] dn + \int_{n_c}^{\infty} n\, P(n,t)\, dn \tag{3.6}$$

remaining constant as a function of time. Note the integration limit at $n = 1$.

As will be seen shortly, the kinetic equations suggest an asymptotic parameterization of the form

$$P_G(n, t) = \zeta(t) c_0 \exp\left[-(\alpha(t))^2 (n - K(t))^2\right], \tag{3.7}$$

for $n > n_c(t)$ and large $t$. We also define the "peak offset"

$$L(t) \equiv n_c(t) - K(t). \tag{3.8}$$

The asymptotic analysis starts with writing (3.5) in a continuous-$n$ form. Unlike Section 2, here we are interested in the precise peak shape and therefore we keep terms up to the second derivative,

$$\frac{\partial P}{\partial t} = (c - c_0)\left[\left(\frac{1}{2}\frac{\partial^2}{\partial n^2} - \frac{\partial}{\partial n}\right)(K_n P)\right]. \tag{3.9}$$

This equation describes the irreversible growth of clusters above the critical size, where, within the assumption of the narrow Gaussian, $P(n,t)$ takes on appreciable values only over a narrow range. Thus we can approximate, for evaluation of the asymptotic behavior, $K_n \approx K_{n_c} = \kappa \left(n_c(t)\right)^{1/3}/c_0$, where $\kappa \equiv 4\pi c_0 a D$. Defining the dimensionless quantity

$$x(t) \equiv c(t)/c_0, \tag{3.10}$$

we get

$$\frac{\partial P}{\partial t} = \kappa (x(t) - 1)\left(n_c(t)\right)^{1/3}\left(\frac{1}{2}\frac{\partial^2}{\partial n^2} - \frac{\partial}{\partial n}\right)P. \tag{3.11}$$

From (3.2), in the asymptotic (large-time) limit $c(t) \to c_0$, we have $x(t) - 1 \approx (2A/3)\left(n_c(t)\right)^{-1/3}$, which cancels the factor $\left(n_c(t)\right)^{1/3}$ in (3.11). For later convenience we introduce the constant $z$ via

$$z^2 \equiv \frac{4A\kappa}{3} = \frac{64\pi^2 a^3 \sigma c_0 D}{3kT}. \tag{3.12}$$

With this definition, (3.11) becomes

$$\frac{\partial P}{\partial t} = \frac{z^2}{2}\left(\frac{1}{2}\frac{\partial^2}{\partial n^2} - \frac{\partial}{\partial n}\right)P. \tag{3.13}$$

Substituting the Gaussian (3.7) into this asymptotic equation for the kinetics, establishes that the solution is indeed of the conjectured form and yields[9] the following asymptotic results for the parameters:

$$\alpha(t) \simeq 1/\sqrt{z^2 t}, \qquad K(t) \simeq z^2 t/2, \qquad \zeta(t) \simeq \Omega/\sqrt{z^2 t}. \tag{3.14}$$

The (constant) coefficient $\Omega$ cannot be determined from the asymptotic analysis alone, because the overall height of the distribution is obviously expected to depend on the initial conditions.

The asymptotic behavior of the peak offset, (3.8), follows from the conservation of matter. Indeed, for large times the second term in (3.6) will be approximated by

$$\int_{n_c(t)}^{\infty} n P_G(n,t) dn, \tag{3.15}$$

which must approach a constant value, equal to the initial total matter less the matter that remains in the thermal distribution as $c \to c_0$. The rather complicated mathematical analysis that follows, will not be reproduced here; see.[9] The key result is that conservation of matter implies

$$L(t) \propto \sqrt{t \ln t} \tag{3.16}$$

for large times. Therefore, the leading asymptotic behavior of the critical cluster size is the same as that for $K(t)$,

$$n_c(t) \simeq z^2 t/2. \tag{3.17}$$

Since the width of the truncated Gaussian is still given by $1/\alpha \sim \sqrt{t}$, we note that our results suggest linear growth of the distribution for large times, see Figure 2 (Page 23), with the *relative width actually decreasing with time*, as $\sim t^{-1/2}$. Finally, one can show[9] that the (positive) difference $c(t) - c_0$ approaches zero as $\sim t^{-1/3}$ for large times.

We comment that the Gaussian distribution has provided[9] a good fit at intermediate and large times for numerical data for various initial conditions, including for initially seeded distributions. Consideration of *seeded distributions* is particularly important because in many actual experiments, especially those involving inorganic solute species, the initial supersaturation is so high that at early times the nucleation barrier is insignificant compared to $kT$, while the calculated values $n_c(t)$ for small times, $t$, will be comparable to or even smaller than $O(1)$. Then the whole "nucleation picture" description cannot be used until fully irreversible fast aggregation processes coarsen the distribution to reduce the supersaturation and build up the nucleation barrier. We can then reset the "initial" time $t = 0$ to that instance, treating the earlier formed cluster-size distribution as the seed for our burst-nucleation approach. Numerical simulations also confirm the other expected features of burst nucleation, summarized in Figure 2 (see Page 23): the initial induction period followed by growth "burst" that precedes the onset of the asymptotically linear growth.

It is experimentally challenging in many situations to unambiguously quantify the size distribution of nucleated nanocrystals, because of their tendency to aggregate, their distribution of non-spherical shapes, and other factors. Still, it is commonly found (and expected) in experiment that the distribution is two-sided around the peak, and that the final particles stop growing after a certain time. Both of these experimental observations are at odds with the predictions of the burst-nucleation model, and the discrepancies can be related primarily to the approximation of instantaneous thermalization of the clusters below the critical size. At very small sizes, below a cutoff value, which can be speculated to correspond to $n_{th} \approx 15\text{-}20$ building blocks[6–8,35–37] (singlets: atoms, molecules, sub-clusters), structures can evolve very rapidly, so that the assumption of fast, thermally driven restructuring is justified.

At larger sizes, however, embryos can be expected to undergo a transition in which their internal atoms assume a more stable, bulk-like crystal structure, and they no longer restructure as easily, except perhaps at their surface layers. Thus for times for which $n_c(t) > n_{th}$, the "classical" nucleation model should be regarded as approximate. Modifications of the model have been contemplated in several previous studies of nucleation.[9,38,39] This, however, requires introduction of new parameters which are not as well defined and as easily experimentally accessible as those of the "classical" nucleation model. In fact, one of the most interesting applications of our present theoretical developments would be to try to estimate, based on experimental data, the deviations from the "classical" behavior and thus obtain information on the value of $n_{th}$, the nanostructure size beyond which a "bulk-material" core develops. A similar effect in colloid synthesis will be mentioned in Section 5.

The extent to which our (unmodified) model describes the initial burst, as well as the range of applicability of the prediction of linear growth of $n_c(t)$, are interesting topics to explore further. We recall that other processes at all cluster sizes, such as cluster-cluster aggregation and ripening, can also modify the kinetics of the distribution, albeit these are usually expected to play role at time scales much larger than the initial nucleation burst.

## 4. UNIFORM POLYCRYSTALLINE COLLOIDS

As described in the preceding section, the burst-nucleation mechanism, which ideally can yield narrow size distributions, is never realized in practice for extended growth times. For larger particles, nucleated in the initial burst and then grown to dimensions typically over several tens of nanometers in diameter, other growth mechanisms usually broaden the size distribution. Here we consider the combined mechanism whereby the nanosize primary particles, burst-nucleated and growing in solution, themselves become the singlets and are "consumed" by the singlet-driven aggregation that results in uniform secondary particles of colloid dimensions. The primary process is thus of the type considered in Section 3, whereas the secondary process is of a variety introduced in Section 2.

A large number of dispersions of uniform colloid particles of various chemical composition and shape, ranging in size from fraction of a micron to few microns, have been synthesized via this route.[1,7,8,18–23,34,40–55] Indeed, it has been found that many spherical particles precipitated from solution showed polycrystalline X-ray characteristics, such as ZnS,[42] CdS,[7,8,41] $Fe_2O_3$,[40] Au and other metals,[1,23,52–54] etc. These particles are not single crystals. Rather, several experimental techniques have confirmed that most monodispersed colloids consist of small crystalline subunits.[1,7,8,18–23,34,40–55] Furthermore, experiments have observed[1,23,50] that the crystalline subunits in the final particles were of the same size as the diameter of the precursor singlets of sizes of order 10 nm, formed in solution, thus suggesting an aggregation-of-subunits mechanism. This two-stage growth process is summarized in Figure 3 (see Page 24). The composite structure has also been identified in uniform non-spherical colloid particles,[40,46–48,55] albeit perhaps thus far not as definitively as for the spherical case.

Let us first outline the simplest "minimal" (in that it avoids introduction of unknown microscopic parameters) model that involves the coupled primary and secondary processes. Even this model requires numerical calculations and cannot be analyzed in closed form. In Section 5, we describe some improvements of the model that allow for better agreement with experimental observations. Additional details, examples of experimental parameters and results, and well as sample numerical data fits can be found in the original works.[1,5,7,8]

The reader might recall that in modeling the burst nucleation process in Section 3, the supercritical distribution was described by the kinetic equations (3.5). However, for the *time dependence* of the subcritical distribution one requires an expression for the time derivative $dc/dt$, since the whole subcritical distribution can be calculated if we know $c(t)$, see (3.3). We did not review, but only referenced our work[9] for mathematical steps, involving the conservation of matter, that give (3.16) and yield an expression for $dc/dt$ (not shown).

When the burst-nucleated supercritical particles are largely consumed by the secondary aggregation, we can assume for simplicity that these primary particles are captured fast enough by the growing secondary particles so that the effect of their aging on the concentration of solutes can be ignored. Furthermore, the radius of the captured primary particles will be assumed close to the critical radius. We discuss the implications of these assumptions later. For now, we write our first expression that applies (approximately) for the two-stage process, but *does not apply* to burst nucleation alone,

$$\frac{dc}{dt} = -n_c \rho . \tag{4.1}$$

Recall that the rate of supercritical particle production, $\rho(t)$, was defined in (3.4) and is a known function of $c(t)$. Thus, this relation, which expresses our approximation that the concentration of solutes is depleted solely due to the irreversible formation of the critical-size nuclei, yields the equations

$$\frac{dc}{dt} = -\frac{2^{14}\pi^5 a^9 \sigma^4 D_a c^2}{(3kT)^4[\ln(c/c_0)]^4} \exp\left\{-\frac{2^8 \pi^3 a^6 \sigma^3}{(3kT)^3[\ln(c/c_0)]^2}\right\}, \tag{4.2}$$

$$\rho(t) = \frac{2^5 \pi^2 a^3 \sigma D_a c^2}{3kT \ln(c/c_0)} \exp\left\{-\frac{2^8 \pi^3 a^6 \sigma^3}{(3kT)^3[\ln(c/c_0)]^2}\right\}. \tag{4.3}$$

These expressions can be used to numerically calculate $\rho(t)$. The notation for various quantities here is the same as in Section 3. However, we denoted by $D_a$ the diffusion constant of the solutes, in order to distinguish it from that of the supercritical primary particles that constitute "singlets" for the secondary process, to be denoted $D_p$.

The growth of the secondary (colloid) particles is facilitated by the appropriate chemical conditions in the system: the ionic strength and/or pH must be kept in ranges such that the surface potential approaches the isoelectric point, resulting in reduction of electrostatic barriers, thus promoting fast irreversible primary particle attachment. Formation of the secondary particles is clearly a diffusion-controlled process.[1,18–23]

We describe the process by the equations for the distribution of growing particles by their size, cf. (2.2)-(2.4), with (2.1). Here it is assumed that the particles are spherical, with the density close to that of the bulk material. Experimentally, the growing particles rapidly restructure to assume the final shape and density: they are not fractal even though the transport of the constituent units is diffusional. The modeling of this restructuring is an interesting unsolved problem on its own, but, as long as the restructuring is fast, its mechanism plays no role in formulating the model equations.

The cluster size $N_{s=1,2,3,...}(t)$ will be defined by how many primary particles (singlets) were aggregated into each secondary particle. The notation here is similar to that in Section 2. For example, relations (2.2)-(2.4), with the notation (2.1), can be solved numerically with the initial conditions $N_{s=1,2,3,...}(0) = 0$. The simplest choice of the rate constants is the Smoluchowski expression

$$\Gamma_s \simeq 4\pi R_p D_p s^{1/3}, \tag{4.4}$$

where $R_p$ is the primary particle radius, and the approximate sign is used because several possible improvement to the simplest formula can be offered, as will be described shortly. A typical numerical calculation for a model of this type is shown in Figure 4 (see Page 25), illustrating the key feature — "size selection" — the "freezing" of the growth even for exponentially increasing times (here ×10).

Let us now discuss some of the numerous simplifying assumptions made in the model just formulated. We will also consider possible modifications of the model. In fact, Figure 4 (see Page 25), which was based on one of the sets of the parameter values used for modeling formation of uniform spherical Au particles, already includes some of the modifications.[5]

We note that since the assumption $s \gg 1$ is not applicable, the full Smoluchowski rate expression[2,27,28] should be used, which, for aggregation of particles of sizes $s_1$ and $s_2$, on encounters due to their diffusional motion, is

$$\Gamma_{s_1,s_2 \to s_1+s_2} \simeq 4\pi \left[ R_p \left( s_1^{1/3} + s_2^{1/3} \right) \right] \left[ D_p \left( s_1^{-1/3} + s_2^{-1/3} \right) \right], \tag{4.5}$$

where for singlet capture $s_1 = s$ and $s_2 = 1$. This relation can not only introduce nontrivial factors for small particle sizes, as compared to (4.4), but it also contains an assumption that the diffusion constant of $s$-singlet, dense particles is inversely proportional to the radius, i.e., to $s^{-1/3}$. This might not be correct for very small, few-singlet aggregates.

Furthermore, another assumption in (4.5), and in (4.4), is that the radius of $s$-singlet, dense particles can be estimated as $R_p s^{1/3}$. However, primary particles actually have a distribution of radii, and they can also age (grow) before their capture by and incorporation into the structure of the secondary particles. In order to partially compensate for the approximations, the following arguments are used.

Regarding the size distribution of the singlets, it can been argued that since their capture rate especially by the larger aggregates is proportional to their radius times their diffusion constant, this rate will not be that sensitive to the particle size and size distribution, because the diffusion constant for each particle is inversely proportional to its radius. Thus, the product is well approximated by a single typical value.

The simplification of ignoring the primary particle ageing, was then further circumvented by using the *experimentally determined* typical primary particle linear size ("diameter"), $2R_{\text{exp}}$, instead of attempting to estimate it as a function of time during the two-stage process. In fact, for the radius of the $s$-singlet particle, the expression in the first square brackets in (4.5), which represents the sum of such terms, $R_p s^{1/3}$, was recalculated with the replacement

$$R_p s^{1/3} \to 1.2 R_{\text{exp}} s^{1/3}. \tag{4.6}$$

Here the added factor is $(0.58)^{-1/3} \simeq 1.2$, where 0.58 is the filling factor of a random loose packing of spheres.[56] It was introduced to approximately account for that as the growing secondary particle compactifies by internal restructuring, not all its volume will be crystalline: a fraction will consists of amorphous "bridging regions" between the nanocrystalline subunits.

A possible inaccuracy in the rate equation (4.1), because primary particles (those not yet captured) age by consuming additional solute matter, which, in fact, can be also directly consumed by the secondary particle surfaces, was partly compensated for[1] by re-normalizing the distribution. This effect seems not to play a significant role in the dynamics. Some additional technical issues and details[1,3,5,7,8] of the modeling are not reviewed here.

Two-stage models of the type just outlined, with *singlet capture* as the main growth mode of the secondary particles, were shown to provide a good semi-quantitative description (without adjustable parameters) of the processes of formation of spherical colloid-size particles of a metal, Au,[1,3,5,7] a salt, CdS,[7,8] as well as argued[57] to qualitatively explain the synthesis of a bio-organic colloid — monodispersed microspheres of Insulin.

# 5. FURTHER DEVELOPMENTS AND DISCUSSION

To advance the agreement between the results of the two-stage model and experimental data for secondary particle size distribution from semi-quantitative to quantitative, additional considerations were required. Here we begin by summarizing these more recent developments, culminating in successful data fits for size distributions of particle of CdS,[7,8] and Au,[58] the former measured for different times during the process and for several protocols of feeding the solutes (which serve as monomers for the primary particle nucleation) into the system, rather than just their instantaneous "batch" supply (which was the case illustrated in Figure 4, see Page 25).

Note that for non-batch supply of atomic-size "monomers," one has to add to the model the rate equations for their production in chemical reactions utilized for their paced release, which is, in itself, an interesting problem, since such reactions, involving the identification and modeling of the kinetics of various possible intermediate solute species, are not always well studied or understood theoretically, and they are not easy to probe experimentally.

The parameters of the primary nucleation process, notably the value of the effective surface tension, were found, by numerical simulations, to mostly affect the time scales of the secondary particle formation, i.e., the onset of "freezing" of their growth as illustrated in Figure 4 (see Page 25). Accumulated evidence suggests that the use of the bulk surface tension and other experimentally determined parameters yields reasonable results consistent with the experimentally observed times.

The kinetic parameters of the secondary process seem to control the average *size* of the final products. We found[1,3,5,7,8,58] that the particle sizes numerically calculated within the "minimal" model, while of the correct order of magnitude, were smaller than the experimentally observed values, by a non-negligible factor. The problem was traced to that the kinetics of the secondary aggregation, as described in Section 4, results in too many secondary particles which, since the total supply of matter is fixed, then grow to sizes smaller than those experimentally observed.

Two explanations for this effect were attempted. The first argued that, especially for very small "secondary" particles, those consisting of one or few primary particles, the spherical-particle diffusional expressions for the rates, which are anyway somewhat ambiguous as described in connection with relations (4.5)-(4.6) vs. (4.4), should be modified. Since the idea is to avoid introduction of many adjustable parameters, the rate $\Gamma_{1,1\to 2}$, cf. (4.5), was modified by a "bottleneck" factor, $f < 1$, with the underlying assumption that "merging" of two singlets (and other very small aggregates) may require substantial restructuring, thus reducing the rate of successful formation of a bi-crystalline entity. The two nanocrystals may instead unbind and diffuse apart, or merge into a single larger nanocrystal, effectively contributing to a new process, $\Gamma_{1,1\to 1}$, not in the original model. However, data fits[5,7] yield values of order $10^{-3}$ or smaller for $f$, which seems too drastic and suggests that alternatives should be seriously considered.

Another modification of the model uses a similar line of argument but in a somewhat different context. We point out that the model already assumes a certain "bottleneck" for particle merger: that of singlet-capture dominance. Indeed, all the rates (4.5) with *both* $s_1 > 1$ and $s_2 > 1$, are set to zero. This assumption was made based on empirical experimental observations that larger particles were never seen to pair-wise "merge" in solution. It seems that the restructuring that leads to rapid compactification of the growing secondary particles, and which is presently not understood experimentally or theoretically, can also "incorporate" primary particles, but not larger aggregates, in the evolving structure, while retaining their crystalline core to yield the final polycrystalline colloids.

One might then argue that perhaps small aggregates, up to certain cutoff sizes, $s_{\max} > 1$, can also be dynamically rapidly incorporated into larger aggregates on diffusional encounters. Thus, we have generalized[7,8] the model equations to allow for cluster-cluster aggregation with rates (4.5), but only as long as *at least one* of the cluster sizes does not exceed $s_{max}$. Now, obviously this sharp cutoff is an approximation, but it offers the convenience of a single new adjustable parameter. Indeed, data fits for CdS and Au spherical particles, yield good quantitative agreement, exemplified in Figure 5 (see Page 26), with fitted values of $s_{max}$ ranging from 15 for Au, to 25 for CdS. Interestingly, these values are not only intuitively reasonable as defining "small" aggregates, but they also fit well with the concept of the cutoff value $n_{th}$, discussed in Section 3, only beyond which atomistic aggregates develop a well formed "bulk-like" core. The only available numerical estimate of such a quantity in solution,[37] for AgBr nano-aggregates, suggests that $n_{th}$ is comparable to or larger than 18 (in terms of molecule count, i.e., the most stable configuration for a $Ag_{18}Br_{18}$ nanocluster is still disordered).

We also comment that cluster-cluster aggregation at small sizes, can explain the formation of the initial peak in the distribution, which later grows by the fast-capture-of-singlets mechanism.

The modification/elaboration of the two-stage model just outlined, required large-scale numerical effort and lead to development of adaptive-mesh (in time and cluster size) algorithmic techniques for efficient simulations.[7,8]

Finally, we point out that in the described treatments we avoided any quantitative or even qualitative modeling of the particle (nanosize and colloid) *shape* selection. Many of the processes that could be treated in a cavalier way in studying the particle sizes will balance to determine the details of the *shape* distribution of the final products. These processes include particle restructuring, both in the interior and at surfaces, as well as monomer transport on particle surfaces and possible monomer detachment/reattachment, as well as detachment/attachment/reattachment/surface motion of larger than monomer structures. The difficulty in modeling these processes is two-fold. Firstly, they are presently not quantified and are difficult to probe experimentally. Secondly, their modeling would require extremely large-scale simulations. Thus, while one can venture guesses as to the key processes that balance to determine the particle shape distribution, derivation of quantitative predictions and their comparison with experimental data remain an important open challenge in colloid and nanoparticles science.

The author gratefully acknowledges collaborations with D. Goia, V. Gorshkov, I. Halaciuga, S. Libert, E. Matijević, D. Mozyrsky, J. Park, D. Robb, and I. Sevonkaev, as well as research funding by the National Science Foundation (grant DMR-0509104).

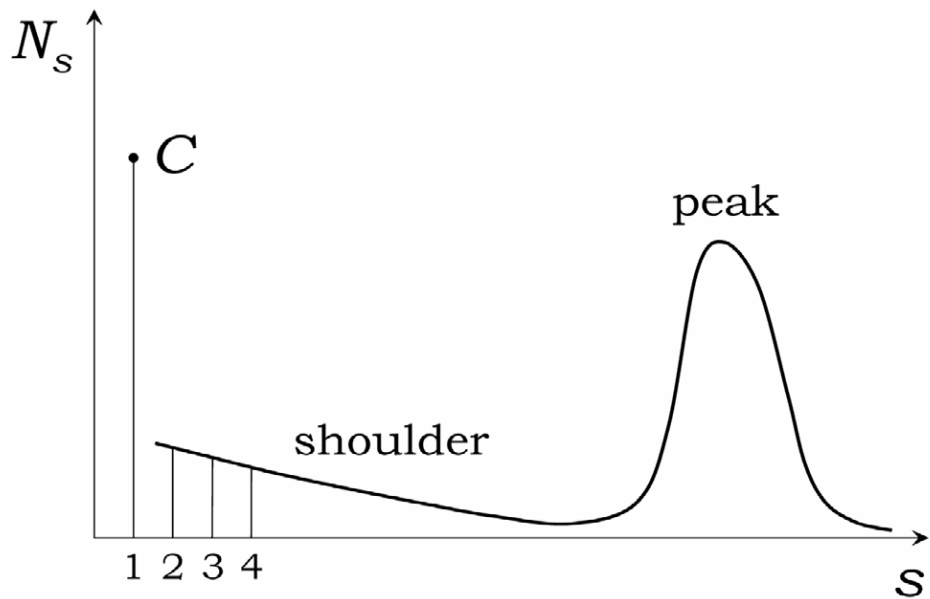

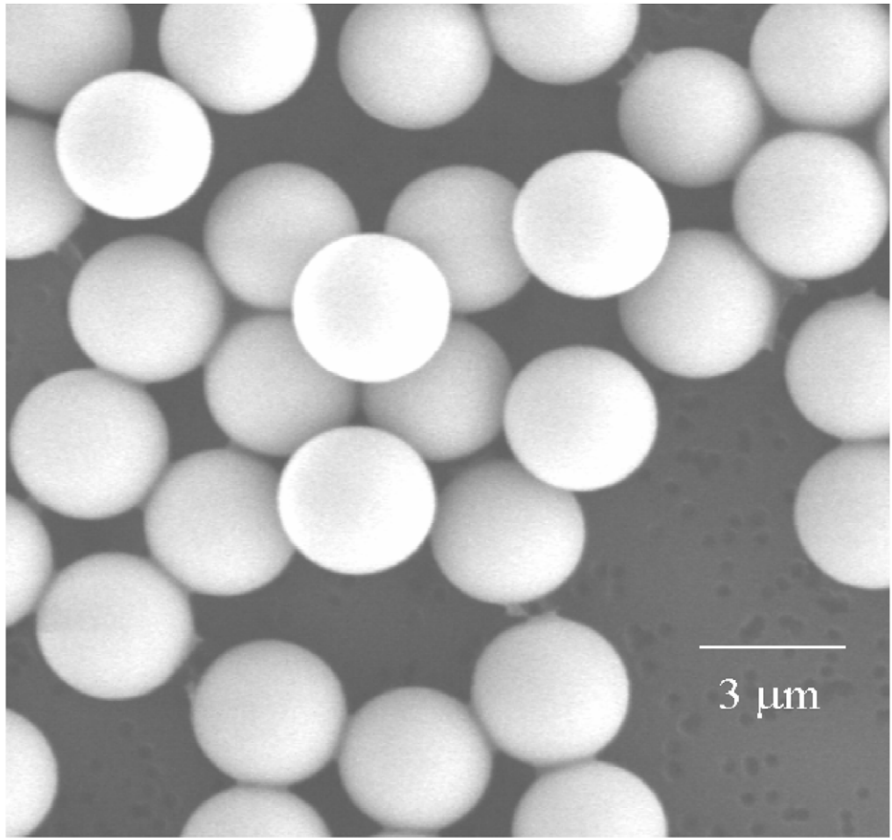

**Figure 1.** Top: The desired particle size distribution. The peak at the larger cluster sizes is growing mostly at the expense of the singlets (which can be supplied externally). The distribution for $s>1$ can be usually assumed a smooth function of $s$, though the vertical bars at $s=1,2,3,4$ emphasize that the $s$ values are actually discrete. Bottom: SEM image of polycrystalline CdS colloid particles illustrating the attainable uniformity of the size and shape distribution.

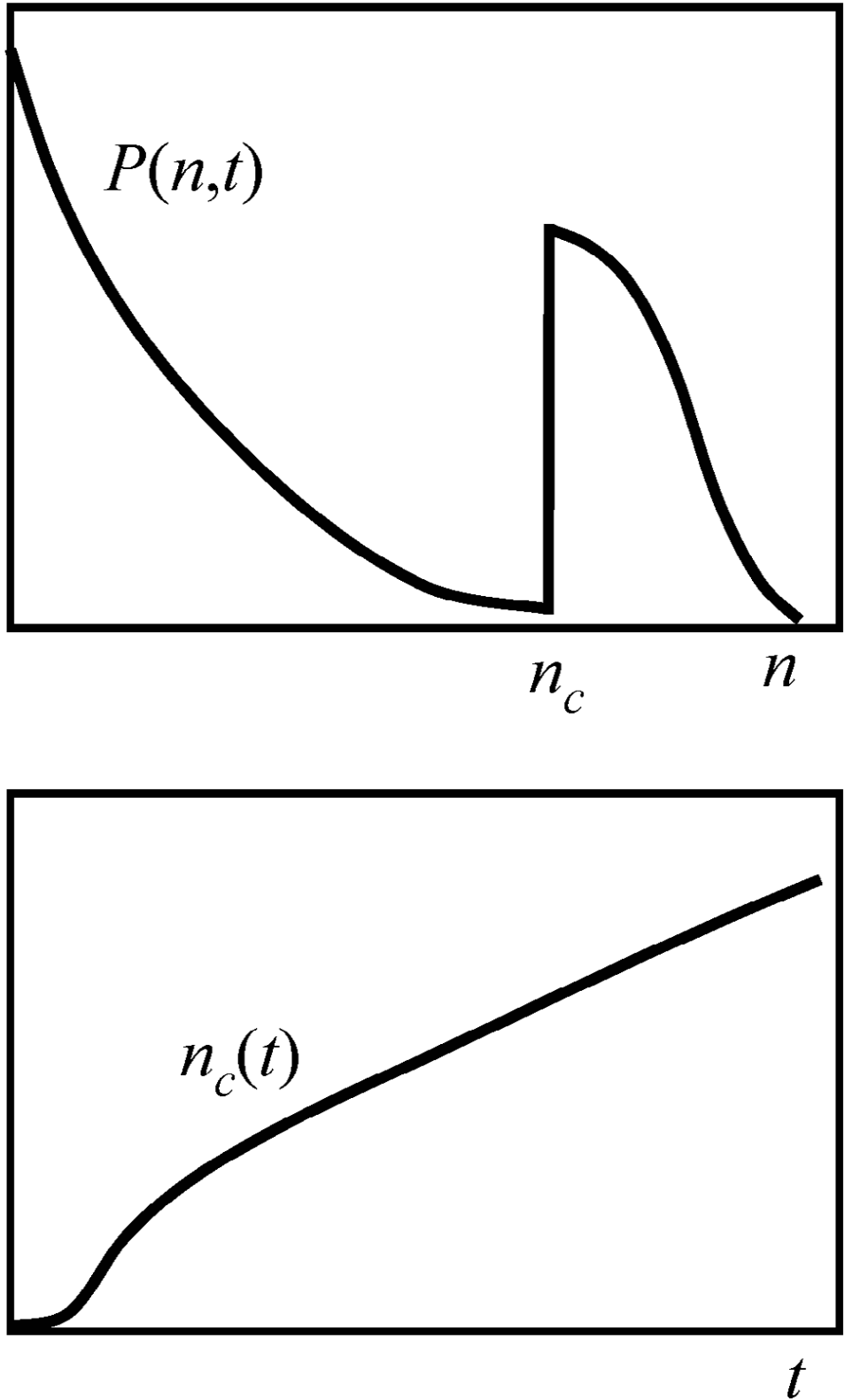

**Figure 2.** Top: Schematic of the asymptotic (large-time) form of cluster size distribution in burst nucleation. Bottom: Schematic of the time dependence of the critical cluster size, showing the initial induction period, followed by the "burst," and then the asymptotically linear growth.

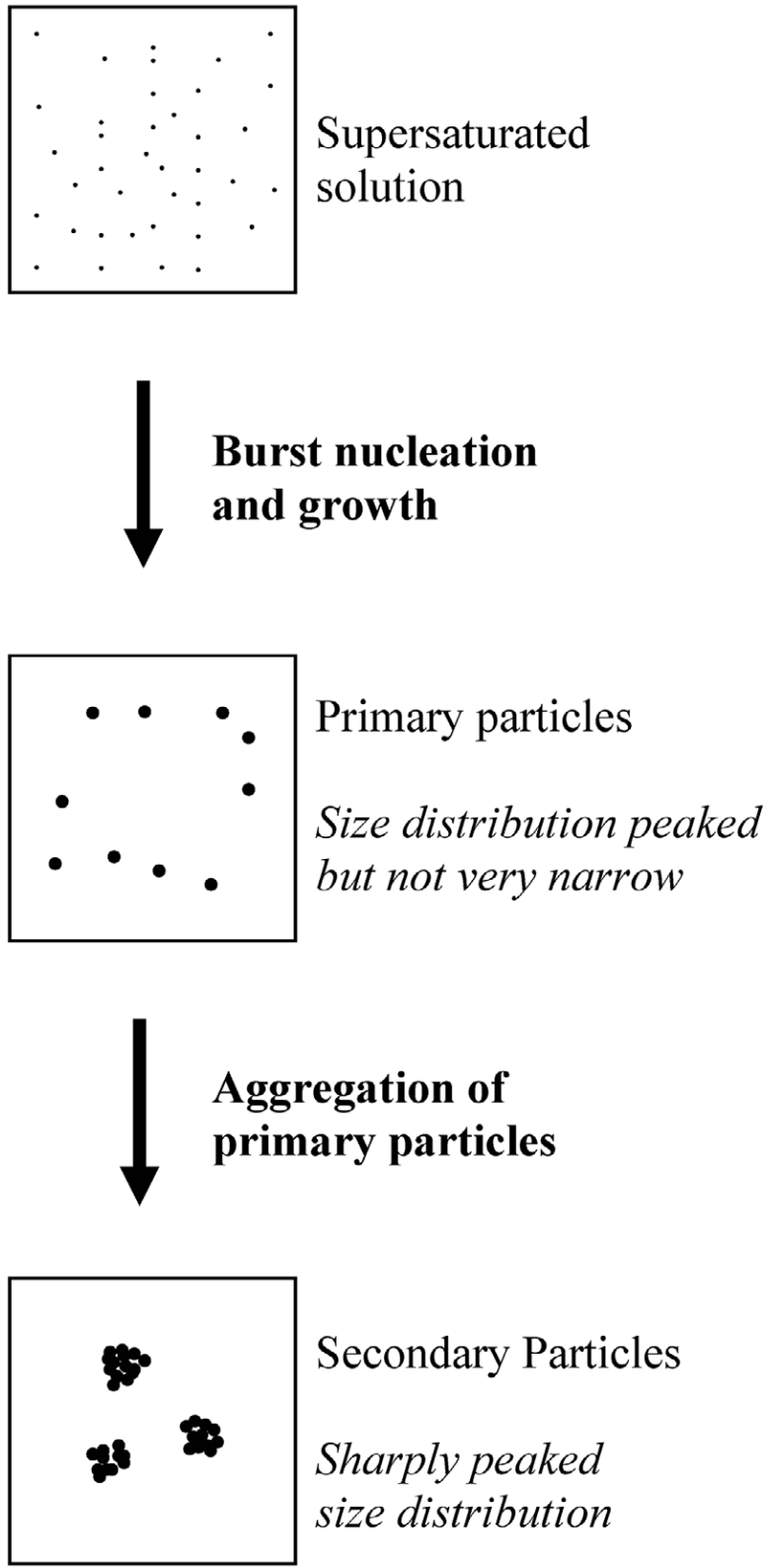

**Figure 3.** Schematic of the two-stage synthesis mechanism of uniform colloids by self-assembly of aggregating nanocrystalline precursor particles which are in turn formed by burst nucleation followed by additional growth that broadens their size distribution.

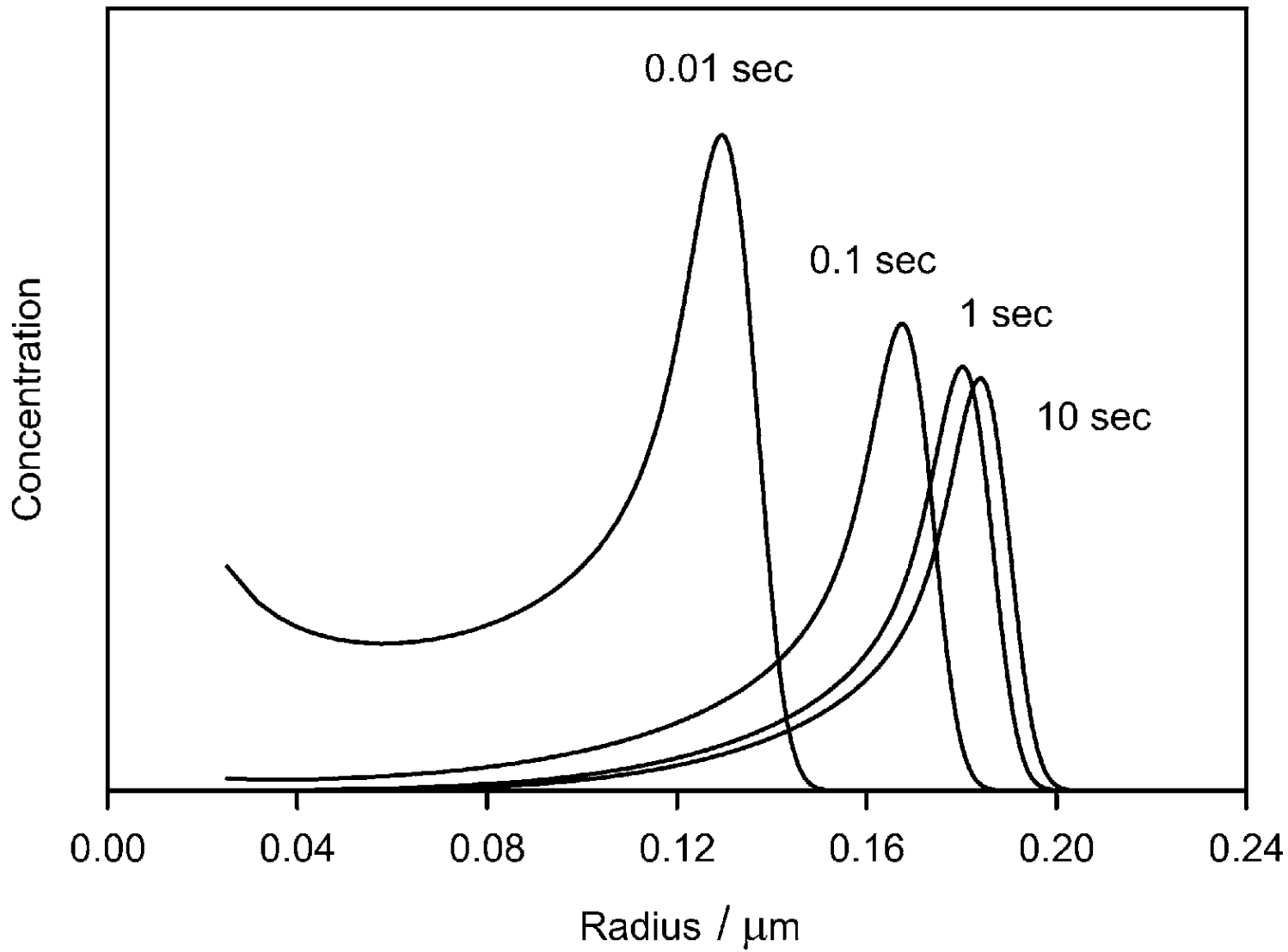

**Figure 4.** Example of a calculated secondary particle size distribution (in arbitrary units), plotted as a function of the colloid particle radius. The parameters correspond to a particular model of formation of spherical Au colloid particles, as referenced in the text.

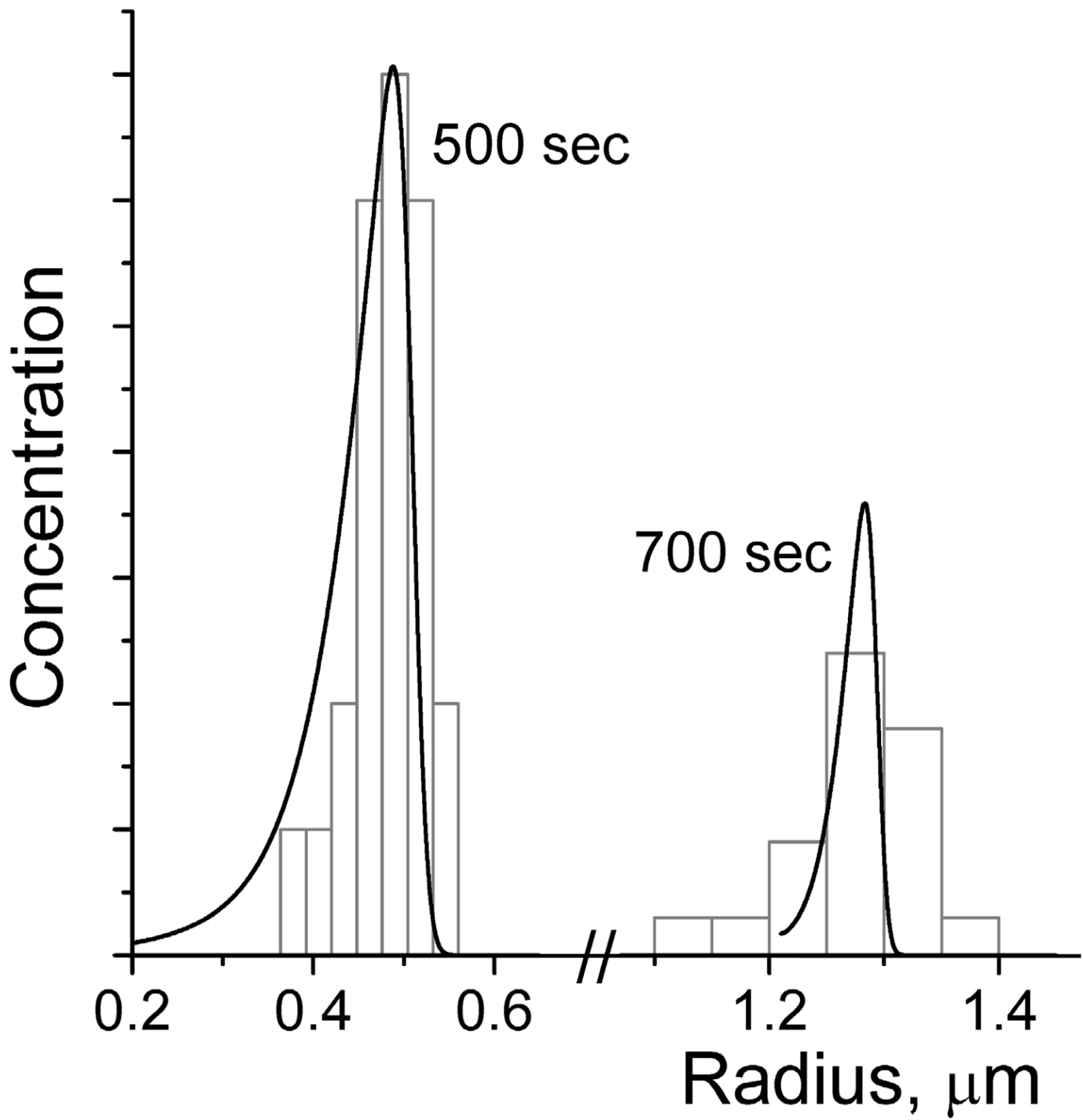

**Figure 5.** Example of a comparison of the calculated (curves) and experimentally measured (histograms) particle size distributions (in arbitrary units), for two different times during the growth, plotted as functions of the particle radius. The parameters correspond to the $s_{\max} = 25$ model of formation of spherical CdS colloids.